\documentclass[useAMS,usenatbib]{mn2e}
\usepackage{times}
\usepackage{epsfig}
\title[A disc model for quasar optical variability]{An accretion disc model for quasar optical variability}
\author[S. Li and X. Cao]{Shuang-Liang
Li\thanks{E-mail:lisl@shao.ac.cn}, Xinwu
Cao\thanks{E-mail:cxw@shao.ac.cn}\\
Shanghai Astronomical Observatory, Chinese Academy of Sciences, 80
Nandan Road, 200030 Shanghai, China}

\begin{document}

\date{Accepted 2008 April 10. Received 2008 April 3}

\pagerange{\pageref{firstpage}--\pageref{lastpage}} \pubyear{2007}

\maketitle

\label{firstpage}

\begin{abstract}

Some different correlations between optical-UV variability and
other quasar properties, such as luminosity, black hole mass and
rest-frame wavelength, were discovered. The positive correlation
between optical-UV variability amplitude and black hole mass was
first found by Wold et al., and this was confirmed by Wilhite et
al. We suggest that the accretion disk model can explain these
correlations, provided the optical-UV variability is triggered by
the change of accretion rate. The disk temperature of accretion
discs decreases with increasing black hole mass, which leads to
systematical spectral shape difference with black hole mass even
if the black hole is accreting at the same rate $\dot{m}$
($\dot{m}=\dot{M}/\dot{M}_{\rm Edd}$). The observed positive
correlation between optical-UV variability and black hole mass can
be well reproduced by our model calculations, if the mean
accretion rate $\dot{m}_0\sim 0.1$ with variation of $\delta
\dot{m}\sim 0.4-0.5 \dot{m}_0$. We also found that the observed
correlations of optical-UV variability amplitude with luminosity
or rest-frame wavelength can be qualitatively explained by this
accretion disc model.

\end{abstract}


\section{Introduction\label{intro}}

Quasars are variable in different wavebands, and their variability
has been extensively explored for several decades
\citep{s1963,m1963}. It has been accepted that quasars are powered
by black hole accretion, and their variability therefore provides an
important clue to understand the physics at work in accretion discs.
The timescales of optical variability for quasars vary from hours to
decades. The variability is found to be correlated with other quasar
properties, such as redshift and luminosity. There is a strong
negative correlation of variability with quasar luminosity, while a
significant positive correlation between variability and redshift is
also present \citep{c1996,vb2004}. An anti-correlation between
variability amplitude and rest-frame wavelength has also been
observed, so the correlation between variability and redshift was
explained by the selection effect, because quasars at higher
redshifts are more easier to be probed in shorter rest-frame
wavelength \citep{cf1996,c1996}. However, \citet{vb2004} claimed
that the evolution of quasar population or the change of quasar
variability mechanism can cause quasars to be more variable at
higher redshifts based on a sample of over 25,000 quasars from the
Sloan Digital Sky Survey, which is different from the selection
effect scenario.

Recently, \citet{wold2007} matched quasars from the the Quasar
Equatorial Survey Team Phase 1 (QUEST1) variability survey with
broad-line objects from the Sloan Digital Sky Survey, which leads to
$\sim100$ quasars. The black hole masses of these quasars are
estimated from the widths of broad-line ${\rm H}_\beta$. They
measured the variability amplitudes from the QUEST1 light curves,
and found a correlation between quasar variability amplitude and
black hole mass.  They found that the correlation seems to be
unlikely caused by selection effects or the contamination coming
from host galaxy or other well-known correlations. This correlation
between black hole mass and variability amplitude was confirmed by
\citet{w2007} by using a sample over 2500 quasars.

There were many efforts devoted to reveal the physics behind the
correlations between quasar optical-UV variability and other
quasar properties. Many different scenarios were suggested, such
as, variation of accretion rates, disc instabilities, stellar
collisions, and microlensing effect
\citep*[e.g.,][]{p2006,m1994,k1998,t1992,t2000,cf2000,m2006}. It
is widely believed that quasars are powered by black hole
accretion. However, the spectral shape in optical-UV bands for a
standard thin accretion disc can be approximated as
$f_\nu\propto\nu^{1/3}$, while the observed quasar continua
usually exhibit $f_\nu\propto\nu^{-1}$. \citet{g2007} suggested
that the observed quasar spectra can be reproduced by accretion
discs with a temperature gradient of $T(R)\propto{R^{-0.57}}$
instead of $T(R)\propto{R^{-0.75}}$ as predicted by standard thin
disc model. This means that there is some flow of heat outwards in
the disk. They use the optical variability timescales of quasars
to argue that the variations must propagate at close to the speed
of light, rather than on viscous timescales \citep{g2007}.
\citet{p2006} found that the multi-epoch quasar spectra in the
rest-frame wavelength range $1300-6000 {\rm \AA}$ can be fitted by
the standard thin disc model, provided their accretion rates vary
from one epoch to the next. Thus, it is a reasonable scenario that
the variability may be caused by the change of accretion rates.

In this work, we investigate the variability amplitude of quasars
as a function of black hole mass based on accretion disc models
assuming the variability to be caused by the change of accretion
rates. We also compare our theoretical calculations with the
observed correlations of variability amplitude with other
quantities (i.e., luminosity or observed wavelength).

\section{Accretion disc models}\label{models}

The big blue bump observed in many AGN is believed to be related
to some kind of thermal emission \citep{m1982,s2002}. The thermal
emission of the standard thin accretion disc can reproduce the
general features of AGN continua, provided some other components,
such as an extrapolation of infrared continuum or a power-law
X-ray continuum, are included \citep{s1989,l1990} \citep*[but also
see][]{g2007}.

The standard thin disc is geometrically thin, optically thick, and
its emission is close to blackbody \citep{s1973}. The gases in the
disc rotate at Keplerian velocities, and they are in hydrodynamical
equilibrium in the vertical direction. The structure of a standard
thin disc can be derived analytically \citep*[see,
e.g.,][]{kato1998}. In this work, we adopt the temperature
distribution of an accretion disc as a function of radius $R$ given
by \citet{kato1998},
\begin{equation}
T(R)=6.9\times10^{7}\alpha^{-1/5}m^{-1/5}\dot{m}^{3/10}{r}^{-3/4}f^{3/10},
\label{tempdisc}
\end{equation}
where, \begin{displaymath} {r}={\frac{R}{R_{\rm g}}},
f\equiv1-\left({\frac {3R_{\rm g}}{R}}\right)^{1/2},
\end{displaymath}
and
\begin{displaymath}
m={\frac {M}{M_\odot}}, \dot{m}={\frac {\dot{M}}{\dot{M}_{\rm
Edd}}}.
\end{displaymath}
Here  $M$ is the black hole mass, $R_{\rm g}=2MG/c^2$, $\dot{M}_{\rm
Edd}=1.5\times10^{18}m{\rm gs^{-1}}, m=M/M_{\odot}$, and $\alpha$ is
the viscosity parameter. The spectrum of the accretion disc can be
calculated with
\begin{equation}
f_\nu=\frac{4{\pi}h\cos{i}\nu^3}{c^2D^2}\int_{R_{\rm in}}^{R_{\rm
out}}\frac{RdR}{e^{h\nu/kT(R)}-1}, \label{spectdisc}
\end{equation}
where $T(R)$ is given by Eq. (\ref{tempdisc}), $R_{\rm in}=3R_{\rm
g}$, $R_{\rm out}$ is the outer radius of disc, $h$ is the Plank's
constant, $i$ is the inclination of axis of the disc with respect to
the line of sight,  and $D$ is the distance from the observer to the
black hole  \citep{f2002}.

The optical/UV continuum of a standard thin disc is
$f_\nu\propto\nu^{1/3}$, which is inconsistent with
$f_\nu\propto\nu^{-1}$ observed in many quasars. \citet{g2007}
suggested that the observed quasar spectra can be reproduced by
accretion discs with a temperature gradient of
$T(R)\propto{R^{-0.57}}$ instead of $T(R)\propto{R^{-0.75}}$ as
predicted by standard thin disc model. Thus, we also adopt this
modified disc model in our spectral calculations (this model is
referred to as modified accretion disc model hereafter). The
temperature of the disc is then given by
\begin{equation}
T(R)={\cal C}_{\rm
cor}6.9\times10^{7}\alpha^{-1/5}m^{-1/5}\dot{m}^{3/10}{r}^{-0.57}f^{3/10},
\label{tempdisc2}
\end{equation}
where ${\cal C}_{\rm cor}$ is a correction factor used to let the
resulted disc luminosity be the same as that for a standard thin
disc with the same disc parameters. The spectra of the modified
accretion discs can be calculated with Eqs. (\ref{spectdisc}) and
(\ref{tempdisc2}).

\section{results}\label{results}

\begin{figure}
\includegraphics[width=8cm]{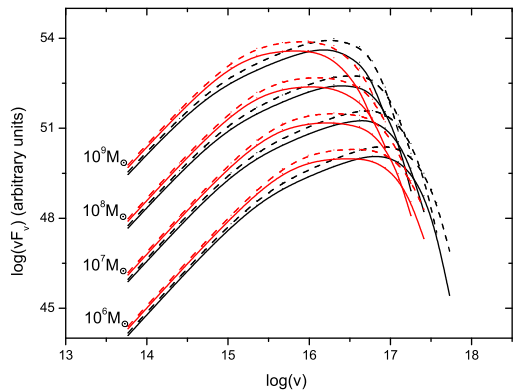}
\caption{The spectra of standard thin discs/modified accretion
discs for different black hole masses. The black and red lines
correspond to standard thin disc model and modified disc model,
respectively. The solid lines correspond to $\dot{m}_0=0.1$, while
the dashed lines represent the cases with
$\dot{m}=\dot{m}_0\pm\delta\dot{m}$
($\delta\dot{m}=0.4\dot{m}_0$). \label{freq}}
\end{figure}

The spectra of standard thin discs can be calculated with Eqs.
(\ref{tempdisc}) and (\ref{spectdisc}), provided the values of
disc parameters ($m$, $\dot{m}$ and $\alpha$) are supplied. In
Fig. \ref{freq}, we plot the spectra of the discs with different
black hole masses and accretion rates. For comparison, we also
plot the spectra of the discs calculated with Eqs.
(\ref{tempdisc2}) and (\ref{spectdisc}) based on the modified disc
model described in last section. In the figure, we find that both
standard thin disc model and modified disc model naturally show
the optical/UV variability becomes larger with the increase of
black hole mass, if the accretion rate changes at the same
percentage.

\begin{figure}
\includegraphics[width=8cm]{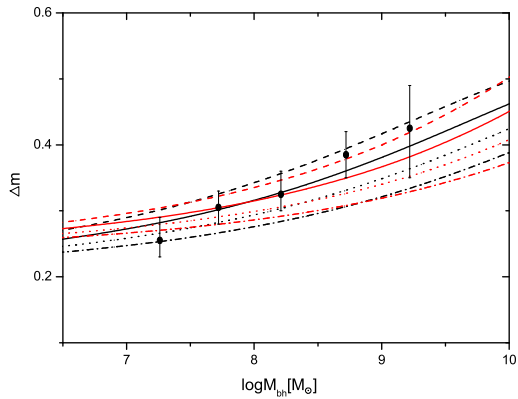}
\caption{The variability amplitude in R-band for accretion discs
as functions of black hole mass.  The black lines and red lines
are for standard thin disc model and modified disc model,
respectively. The dashed, solid, dotted and dashed-dotted line are
for accretion rates $\dot m_0$=0.05, 0.1, 0.2 and 0.4
respectively. All the calculations are carried out by assuming
accretion rates to vary with $\delta \dot{m}\sim 0.4 \dot{m}_0$
for standard thin disc model ($\delta \dot{m}\sim 0.5 \dot{m}_0$
for the modified thin disc model). The five points with error bars
are the statistical results given in Fig. 5 of Wold et al. (2007)
(see their paper for the details).  The solid lines can fit the
statistical results quite well. \label{varia}}
\end{figure}

Assuming the optical variability to be caused by the change of
accretion rates in AGN, we calculate the optical variability
amplitude as a function of black hole mass with the disc models
described in last section. In our calculations, the accretion rate
varies in the range of $\dot{m}=\dot{m}_0\pm\delta \dot{m}$, where
$\dot{m}_0=0.1$ and $\delta\dot{m}=0.4\dot{m}_0$ for standard disc
model ($\delta\dot{m}=0.5\dot{m}_0$ for the modified disc model),
are adopted. In Fig. \ref{varia}, we compare our calculations with
the correlation between variability amplitude at R-band and black
hole mass discovered by \citet{wold2007}. We find that our
calculations can reproduce their statistic result quite well(see
Fig. \ref{varia}).

\begin{figure}
\includegraphics[width=8cm]{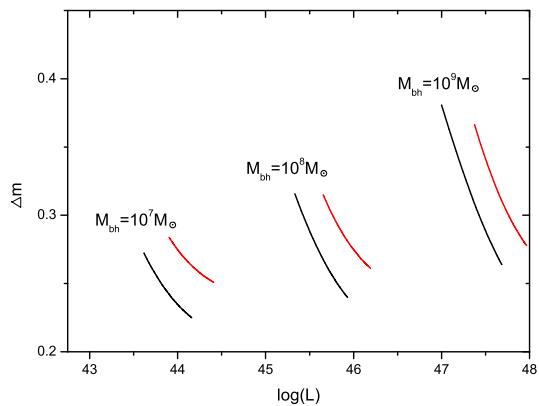}
\caption{The variability amplitude as functions of luminosity for
different black hole masses. The black lines and red lines are for
standard thin disc model and the modified disc model,
respectively, which are the same as in Fig. \ref{freq}.
\label{lumin}}
\end{figure}

We also calculate the optical variability amplitude as functions
of luminosity by changing accretion rates $\dot{m}$ for given
black hole mass $m$ (see Fig. \ref{lumin}). We find that the
optical variability amplitude decreases with increasing luminosity
for a given black hole mass, which is qualitatively consistent
with the observed anti-correlation between these two quantities
\citep*[e.g.,][]{c1996,vb2004,w2007}.

\begin{figure}
\includegraphics[width=8cm]{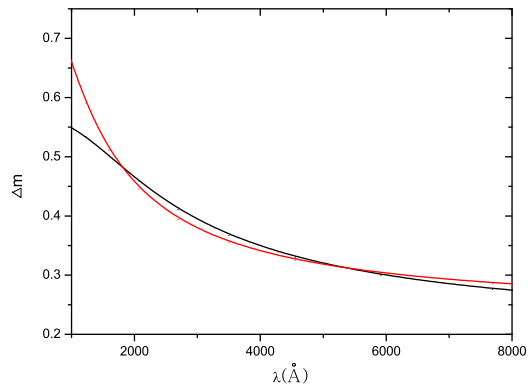}
\caption{The variability amplitude of different accretion rate
changes as a function of rest-frame wavelength. The arrangement of
black line, red line, accretion rate and variation of accretion
rate are the same with Fig. \ref{freq}. \label{wavelength}}
\end{figure}

The dependence of the optical variability amplitude on the
observed wavelength measured in the rest frame can be calculated
based on either the standard thin disc model or the modified disc
model for given disc parameters (see Fig. \ref{wavelength}). It is
found that the variability amplitude decreases with increasing
rest-frame wavelength is either for standard accretion disc model
or the modified disc model.

\section{conclusions and discussion}\label{conclusions and discussion}

The temperature of the disc decreases with increasing black hole
mass, so the spectra of the discs surrounding massive black holes
are softer than those for less massive black holes (see Fig.
\ref{freq}). It is found that the optical variability amplitude
varies with black hole mass even if the changes of accretion rates
$\dot{m}$ are same. Our present calculations show that the observed
correlation between the optical variability amplitude and black hole
mass can be explained by accretion disc models, if the variability
is triggered by the variation of accretion rate. In all our
calculations, a conventional value of viscosity parameter
$\alpha=0.1$ is adopted. The final results depend on the value of
$\alpha$ quite insensitively (see Eqs. \ref{tempdisc} and
\ref{tempdisc2}), and the conclusions will not be altered even if a
different value of $\alpha$ is adopted.

In this work, the accretion rate $\dot{m}=0.1\pm 0.04$ (standard
thin disc model) and $\dot{m}=0.1\pm 0.05$ (modified disc model)
are adopted to calculate the disc spectra, and we find that the
optical variability amplitude increases with black hole mass,
which can reproduce the observed correlation quite well (see Fig.
\ref{varia}). The mean accretion rate of the sample used in
\citet{wold2007} is $\simeq 0.0955$, which is roughly consistent
with the mean accretion rate $\dot{m}=0.1$ adopted in this work
provided the radiative efficiency is $\sim 0.1$.


The optical emission is dominantly from the accretion disc region
with several hundred Schwarzschild radii from the black hole
\citep*[e.g.,][]{g2007,l2008}. The timescales of quasar
variability in \citet{wold2007} range from a few hours to years,
which are much shorter than the viscous timescale of the disc, but
is usually longer than the light-travel time of the distance to
the black hole \citep*[see detailed discussion in][]{g2007}. This
may imply that the energy propagates at close to the speed of
light, rather than on viscous timescales. Thus, the outer disc
region is irradiated by the radiation of the inner disc region,
and the temperature distribution of the disc is altered
significantly as described by the modified accretion disc model.
Such a modified disc model can successfully reproduce the steep
spectra in optical-UV bands \citep{g2007}.

Our calculations indicate that the optical variability amplitude
declines with luminosity of the accretion disc if it black hole
mass is fixed. This is qualitatively  consistent with the
statistical result given by \citet{w2007}. Based on either the
standard accretion disc model or the modified disc model, the
spectral calculation show that the optical variability amplitude
decreases with increasing wavelength. Such a relation has already
observed by \citet{vb2004}.

We thank Zhaohui Shang for helpful discussion and the support from
the NSFC (grant 10773020), the CAS (grant KJCX2-YW-T03) and the
China Postdoctoral Science Foundation (grant 20070420681).

\label{lastpage}

\end{document}